\documentclass{article}
\newcommand {\Pbar}{{\mbox{\rm$\mbox{I}\!\mbox{P}$}}}
\newcommand  {\Ebar} {{\mbox{\rm$\mbox{I}\!\mbox{E}$}}}
\newcommand  {\Rbar} {{\mbox{\rm$\mbox{I}\!\mbox{R}$}}}

\newcommand  {\Nbar} {{\mbox{\rm$\mbox{I}\!\mbox{N}$}}}
\newcommand {\Tbar}{{\mbox{\rm$\mbox{I}\!\!\!\mbox{T}$}}}

\newsavebox{\zzzbar}
\sbox{\zzzbar}
  {\setlength{\unitlength}{0.9em}
  \begin{picture}(0.6,0.7)
  \thinlines
  \put(0,0){\line(1,0){0.6}}
  \put(0,0.75){\line(1,0){0.575}}
  \multiput(0,0)(0.0125,0.025){30}{\rule{0.3pt}{0.3pt}}
  \multiput(0.2,0)(0.0125,0.025){30}{\rule{0.3pt}{0.3pt}}
  \put(0,0.75){\line(0,-1){0.15}}
  \put(0.015,0.75){\line(0,-1){0.1}}
  \put(0.03,0.75){\line(0,-1){0.075}}
  \put(0.045,0.75){\line(0,-1){0.05}}
  \put(0.05,0.75){\line(0,-1){0.025}}
  \put(0.6,0){\line(0,1){0.15}}
  \put(0.585,0){\line(0,1){0.1}}
  \put(0.57,0){\line(0,1){0.075}}
  \put(0.555,0){\line(0,1){0.05}}
  \put(0.55,0){\line(0,1){0.025}}
  \end{picture}}
\newcommand{\Zbar}{\mathord{\!{\usebox{\zzzbar}}}}

\newsavebox{\uuunit}
\sbox{\uuunit}
    {\setlength{\unitlength}{0.825em}
     \begin{picture}(0.6,0.7)
        \thinlines
        \put(0,0){\line(1,0){0.5}}
        \put(0.15,0){\line(0,1){0.7}}
        \put(0.35,0){\line(0,1){0.8}}
       \multiput(0.3,0.8)(-0.04,-0.02){12}{\rule{0.5pt}{0.5pt}}
     \end {picture}}

\newcommand{\QED}{{\hspace*{\fill}\rule{2mm}{2mm}\linebreak}}

\newtheorem{lemma}{Lemma}[section]
\newtheorem{proposition}{Proposition}[section]
\newtheorem{theorem}{Theorem}[section]
\newtheorem{definition}{Definition}[section]

\newcommand{\Z}{\Zbar}
\newcommand{\R}{\Rbar}
\newcommand{\N}{\Nbar}

\newcommand{\E}{\Ebar}
\newcommand{\T}{\Tbar}
\newcommand{\pee}{\Pbar}
\newcommand{\bee}{\ensuremath{\mathcal{B} }}
\newcommand{\re}{\ensuremath{\mathcal{R}}}
\newcommand{\ce}{\ensuremath{\mathcal{C}}}

\newcommand{\na}{\ensuremath{N^{t,x}_\varphi}}
\newcommand{\aaa}{\ensuremath{\mathcal{A}}}
\newcommand{\vi}{\ensuremath{\varphi}}
\begin{document}
\pagestyle{myheadings} \markright{Abelian Sandpile Model}
\setlength{\textheight}{21cm}
\title{{\bf The Abelian Sandpile Model on an Infinite Tree}
}
\author{{\bf Christian Maes, {\it K.U.Leuven}} \\
{\bf Frank Redig, {\it T.U. Eindhoven}}\\
{\bf Ellen Saada, {\it C.N.R.S., Rouen}}} \maketitle
 \footnotesize
\begin{quote}
{\bf Abstract:} We consider the standard Abelian sandpile process on the
Bethe lattice. We show the existence of the thermodynamic limit for the
finite volume stationary measures and the existence of a unique infinite
volume Markov process exhibiting features of self-organized
criticality.

\end{quote}
\normalsize

\vspace{12pt}

\section{Introduction}
Global Markov processes for spatially extended systems have been around
for about 30 years now and interacting particle systems have become a
branch of probability theory with an increasing number of connections
with the natural and human sciences.  While standard techniques and
general results have been collected in a number of books such as
\cite{Ligg2,Mu,TVSMKP} and are capable to treat the infinite volume
construction for stochastic systems with locally interacting components,
some of the most elementary questions for long range and nonlocal
dynamics have remained wide open.  We have in mind the class of
stochastic interacting systems that during the last decade have invaded
the soft condensed matter literature and are sometimes placed under the
common denominator of self-organizing systems.

Since the appearance of the paper \cite{Bak}, the concept of
self-organized criticality (SOC) has suscited much interest, and is
applied in a great variety of domains (see e.g. \cite{turcotte} for an
overview). From the mathematical point of view, the situation is however
quite unsatisfactory. The models exhibiting SOC are in general very
boundary condition dependent (especially the BTW model in dimension 2),
which suggests that the definition of an infinite volume dynamics poses a
serious problem. Even the existence of a (unique) thermodynamic limit of
the finite volume stationary measure is not clear. From the point of view
of interacting particle systems no standard theorems are at our disposal.
The infinite volume processes we are looking for will be non-Feller and
cannot be constructed by monotonicity arguments as in the case of the
one-dimensional BTW model (see \cite{MRV}) or the long-range exclusion
process (see \cite{Ligg1}). On the other hand in order to make
mathematically exact statements about SOC, it is necessary to have some
kind of infinite volume limit, both for statics and for dynamics.

In this paper we continue our study of the BTW-model for the case of the
Bethe lattice, this is the abelian sandpile model on an infinite tree.
For this system, many exact results were obtained in \cite{DharMajum}. In
contrast to the one-dimensional case this system has a non-trivial
stationary measure. We show here that the finite volume stationary
measures converge to a unique measure $\mu$ which is not Dirac and
exhibits all the properties of a SOC-state. We then turn to the
construction of a {\it stationary} Markov process starting from this
measure $\mu$. The main difficulty to overcome is the strong non-locality:
Adding a grain at some lattice site $x$ can influence the configuration
far from $x$. In fact the cluster of sites influenced by adding at some
fixed site has to be thought of as a critical percolation cluster which
is almost surely finite but not of integrable size. The process we
construct is intuitively described as follows: At each site $x$ of the
Bethe lattice we have an exponential clock which rings at rate $\vi (x)$.
At the ringing of the clock we add a grain at $x$. Depending on the
addition rate $\vi (x)$, we show existence of a stationary Markov process
which corresponds to this description. We also extend this stationary
dynamics to initial configurations which are typical for a measure $\mu'$
that is stochastically below $\mu$.

The paper is organized as follows. In section 2 we introduce standard
results on finite volume abelian sandpile models, and summarize some
specific results of \cite{DharMajum} for the Bethe lattice which we need
for the infinite volume construction. In section 3 we present the results
on the thermodynamic limit of the finite volume stationary measures and
on the existence of infinite volume Markovian dynamics. Section 4 is
devoted to proofs and contains some additional remarks.

\section{Finite Volume Abelian Sandpiles}
In this section we collect some results on abelian sandpiles on finite
graphs which we will need later on.  Most of these results are contained
in the review paper \cite{Dhar1}, or in \cite{IP}.

\subsection{Toppling Matrix}
Let $V$ denote a finite set of sites and
$\Delta^V=(\Delta_{x,y}^V)_{x,y\in V}$ a matrix indexed by the elements
of $V$ satisfying the conditions :
\begin{enumerate}
\item
For all $x,y\in V$, $x\not=y$, $\Delta_{x,y}^V=\Delta_{y,x}^V\leq 0$,
\item
For all $x\in V$, $\Delta_{x,x}^V\geq 1$,
\item
For all $x\in V$, $\sum_{y\in V} \Delta_{x,y}^V \geq 0$,
\item
$\sum_{x,y\in V} \Delta_{x,y}^V > 0.$
\end{enumerate}
Such a matrix $\Delta^V$ is called a {\sl toppling matrix}. The fourth
condition ensures that there are sites (so called {\sl dissipative
sites}) for which the inequality in the third condition is strict. This
is fundamental for having a well defined toppling rule later on. In the
rest of the paper we will choose $\Delta^V$ to be the lattice Laplacian
with open boundary conditions on a finite simply-connected set $V\subset
S$, where $S$ is a regular graph, like the $d$-dimensional lattice $\Z^d$,
or the infinite rootless tree $\T_d$ of degree $d+1$. More explicitly:
\begin{eqnarray}\label{laplace}
\Delta_{x,x}^V&=& 2d \ \mbox{ if}\ V\subset \Z^d,\nonumber\\
&=& d+1 \ \mbox{if }V  \subset \T_d,\nonumber\\
\Delta_{x,y}^V&=& -1 \ \mbox{ if}\ x\  \mbox{and}  \ y \ \mbox{are
nearest  neighbors}.
\end{eqnarray}
The dissipative sites then correspond to the boundary sites of $V$.

\subsection{Configurations}
A {\sl height configuration} $\eta$ is a mapping from $V$ to
$\N=\{1,2,...\}$ assigning to each site a natural number $\eta (x) \geq 1$
(``the number of sand grains" at site $x$). A configuration $\eta \in
\N^V$ is called {\sl stable} if, for all $x\in V$, $\eta(x)\leq
\Delta_{x,x}^V$. Otherwise $\eta$ is {\sl unstable}. We denote by
$\Omega_V$ the set of all stable height configurations. For  $\eta \in
\N^V$ and $V'\subset V$, $\eta|_{V'}$ denotes the restriction of $\eta$
to $V'$.

\subsection{Toppling Rule}
The {\sl toppling rule } corresponding to the toppling matrix $\Delta^V$
is the mapping
\[
T_{\Delta^V}: \N^V\times V \rightarrow \N^V
\]
defined by
\begin{eqnarray}\label{Tdelta}
T_{\Delta^V }(\eta,x ) (y) &=& \eta (y) -\Delta_{x,y}^V \ \mbox{if}
\ \eta(x) > \Delta_{x,x}^V,\nonumber\\
&=& \eta (y) \ \mbox{otherwise}.
\end{eqnarray}
In words, site $x$ topples if and only if its height is strictly larger
than $\Delta_{x,x}^V$, by transferring $-\Delta_{x,y}^V$ grains to site
$y\not= x$ and losing itself $\Delta_{x,x}^V$ grains. Toppling rules
commute on unstable configurations. This means for $x,z\in V$ and $\eta$
such that $\eta(x) > \Delta_{x,x}^V,\eta(z) > \Delta_{z,z}^V$,
\begin{equation}\label{abrule}
T_{\Delta^V }\left(T_{\Delta^V }(\eta,x ),z\right)= T_{\Delta^V }
\left(T_{\Delta^V }(\eta,z ),x\right)
\end{equation}
We write $[T_{\Delta^V }(\cdot,z )T_{\Delta^V }(\cdot,x )](\eta)=
[T_{\Delta^V }(\cdot,x )T_{\Delta^V }(\cdot,z )](\eta)$.

Choose some enumeration $\{ x_1,\cdots,x_n \}$ of the set $V$. The {\sl
toppling transformation} is the mapping
\[
\mathcal{T}_{\Delta^V}: \N^V\rightarrow \Omega_V
\]
defined by
\begin{equation}\label{deftop}
\mathcal{T}_{\Delta^V}(\eta ) = \lim_{N\uparrow\infty} \left(
\prod_{i=1}^n T_{\Delta^V}(\cdot,x_i) \right)^N (\eta ).
\end{equation}
In \cite{IP} it is recalled that
\begin{enumerate}
\item
The limit in (\ref{deftop}) exists, i.e. there are no infinite cycles,
due to the presence of dissipative sites.
\item
The stable configuration $\mathcal{T}_{\Delta^V }(\eta)$ is independent
of the chosen enumeration of $V$. This is the {\sl abelian property} and
follows from (\ref{abrule}).
\end{enumerate}

\subsection{Addition Operators}
For $\eta\in \N^V$ and $x\in V$, let $\eta^x$ denote the configuration
obtained from $\eta$ by adding one grain to site $x$, i.e. $\eta^x
(y)=\eta (y) +\delta_{x,y}$. The {\sl addition operator} defined by
\begin{equation}\label{axV}
a_{x,V}:\Omega_V\rightarrow\Omega_V;\eta\mapsto a_{x,V}\eta =
\mathcal{T}_{\Delta^V} (\eta^x)
\end{equation}
represents the effect of adding a grain to the stable configuration
$\eta$ and letting the system topple until a new stable configuration is
obtained. By (\ref{abrule}), the composition of addition operators is
commutative: For all $\eta \in \Omega_V,\,x,y\in V$,
\[
a_{x,V} (a_{y,V}\eta ) = a_{y,V} (a_{x,V} \eta ).
\]

\subsection{Finite Volume Dynamics}
Let $p$ denote a non-degenerate probability measure on $V$, i.e. numbers
$p_x$, $0< p_x< 1$ with $\sum_{x\in V} p_x =1$. We define a discrete time
Markov chain $\{\eta_n: n\geq 0\}$ on $\Omega_V$ by picking a point $x\in
V$ according to $p$ at each discrete time step and applying the addition
operator $a_{x,V}$ to the configuration. This Markov chain has the
transition operator
\begin{equation}\label{2.9}
P_V f(\eta ) = \sum_{x\in V}p_x f(a_{x,V}\eta ).
\end{equation}
We can equally define a continuous time Markov process $\{\eta_t : t\geq
0 \}$ with infinitesimal generator
\begin{equation}\label{2.10}
L_V^\vi f(\eta ) = \sum_{x\in V} \varphi(x)[ f(a_{x,V}\eta ) -f (\eta )],
\end{equation}
generating a pure jump process on $\Omega_V$, with addition rate
$\varphi(x)>0$ at site $x$.

\subsection{Recurrent Configurations, Stationary Measure}
We see here that the Markov chain $\{\eta_n, n\geq 0\}$ has only one
recurrent class and its stationary measure is the uniform measure on that
class.

Let us call $\mathcal{R}_V$ the set of recurrent configurations  for
$\{\eta_n, n\geq 0 \}$, i.e. those for which $P_\eta (\eta_n = \eta \
\mbox{infinitely often}) = 1,$ where $P_\eta$ denotes the distribution of
$\{ \eta_n, n\geq 0 \}$ starting from $\eta_0 =\eta\in \Omega_V$. In the
following proposition we list some properties of $\re_V$. For the sake of
completeness we include a proof which we could not find worked out
completely in the literature.

\begin{proposition}
\noindent
\begin{enumerate}
\item $\re_V$ contains only one recurrent class.
\item The composition of the addition operators $a_{x,V}$ restricted to
$\re_V$ defines an abelian group $G$.
\item $|G|=|\re_V|$.
\item For any $x\in V$, there exists $n_x$
such that for any $\eta\in\re_V$, $a_{x,V}^{n_x}\eta=\eta$.
\item $|\re_V|=\mbox{\rm det } \Delta^V$.
\end{enumerate}
\end{proposition}

\noindent{\bf Proof:} 1. We write $\eta\hookrightarrow \zeta$ if in the
Markov chain $\zeta$ can be reached from $\eta$ with positive
probability. Since sand is added with positive probability on all sites
($p_x>0$), the {\sl maximal configuration} $\eta_{max}$ defined by
\[
\eta_{max} (x)= \Delta_{x,x}^V
\]
can be reached from any other configuration. Hence, if $\eta\in\re_V$
then $\eta\hookrightarrow \eta_{max}$, therefore $\eta_{max}\in\re_V$ and
$\eta_{max}\hookrightarrow \eta$ (see e.g. \cite{Chung} p.19).

2. Fix $\eta\in\re_V$; then there exist $n_y\geq 1$ such that
\[
\prod_{y\in V} a_{y,V}^{n_y} \eta =\eta,
\]
and
\[
g_x=a_{x,V}^{n_x-1}\prod_{y\in V, y\not=x}a_{y,V}^{n_y}
\]
satisfies $(a_{x,V} g_x)(\eta ) = (g_x a_{x,V})( \eta) =\eta$. The set
\[
\re^x=\{\zeta\in\re_V:(a_{x,V} g_x) (\zeta) =\zeta \}
\]
is closed under the action of $a_{x,V}$, contains $\eta$, hence also
$\eta_{max}$: it is a recurrent class. By  part 1, $\re^x=\re_V$,
$a_{x,V}g_x$ is the neutral element $e$, and $g_x=a_{x,V}^{-1}$ if we
restrict $a_{x,V}$ to $\re_V$.

3. Fix $\zeta\in\re_V$ and put $\Psi_\zeta : G\rightarrow\re_V; g\mapsto
g(\zeta).$ As before $\Psi_\zeta (G)$ is a recurrent class, hence
$\Psi_\zeta (G) = \re_V$. If for $g,h\in G$, $\Psi_\zeta (g) = \Psi_\zeta
(h)$, then $gh^{-1} (\zeta) = \zeta$, and by commutativity $gh^{-1}
(g'\zeta ) = g'\zeta$ for any $g'\in G$. Therefore $gh^{-1}(\xi)=\xi$ for
all $\xi\in \re_V$, thus $g=h$. This proves that $\Psi_\zeta$ is a
bijection from $G$ to $\re_V$.

4. Since $G$ is a finite group, for any $x\in V$ there exists $n_x\geq 1$
such that $a_{x,V}^{n_x} =e$.

5. Adding $\Delta_{x,x}^V$ particles at a site $x\in V$ makes the site
topple, and $-\Delta_{x,y}^V$ particles are transferred to $y$. This gives
\[
a_{x,V}^{\Delta_{x,x}^V} = \prod_{y\not=x} a_{y,V}^{-\Delta_{x,y}^V}.
\]
On $\re_V$ the $a_{x,V}$ can be inverted and we obtain the closure
relation
\[
\prod_{y\in V}a_{y,V}^{\Delta_{x,y}^V} = e,
\]
which completely determines the one-dimensional representations of the
group of addition operators, and in particular the cardinality of the
latter, as obtained in \cite{Dhar}. \QED

\noindent {\bf Remark.} $\re_V$ does not depend on the $p_x$, and does
not change by going from discrete to continuous time, i.e. from
(\ref{2.9}) to (\ref{2.10}).

The main consequence of the group property of $G$ is the fact that the
unique stationary measure is uniform on $\re_V$.

\begin{proposition}
\noindent
\begin{enumerate}
\item
The measure
\begin{equation}\label{finitevolume}
\mu_V = \sum_{\eta\in \re_V} \frac{1}{|\re_V|} \delta_\eta
\end{equation}
is invariant under the action of $a_{x,V}$, $x\in V$ ($\delta_\eta$ is the
Dirac measure on configuration $\eta$).
\item
On $L^2(\mu_V)$ the adjoint of $a_{x,V}$ is
\begin{equation}\label{adjointtrans}
a^*_{x,V} =a_{x,V}^{-1}.
\end{equation}
\end{enumerate}
\end{proposition}

\noindent{\bf Proof:} Since  $a_{x,V}:\re_V\rightarrow\re_V$ can be
inverted, we have
\[
\sum_{\eta\in \re_V} f(a_{x,V}\eta) g(\eta ) = \sum_{\eta\in \re_V}
f(\eta ) g(a_{x,V}^{-1} \eta ),
\]
hence (\ref{adjointtrans}). By choosing $g\equiv 1$, part 1 follows. \QED

\noindent {\bf Remark.} This shows that $\mu_V$ is invariant under the
Markov processes generated by (\ref{2.9}) and (\ref{2.10}).

\subsection{Burning Algorithm}
The {\sl burning algorithm} determines whether a stable configuration
$\eta\in \Omega_V$ is recurrent or not. It is described as follows: Pick
$\eta\in \Omega_V$ and erase all sites $x\in V$ satisfying the inequality
\[
\eta (x) > \sum_{y\in V,y\not= x} (-\Delta_{x,y}^V).
\]
This means ``erase the set $E_1$ of all sites $x\in V$ with a height
strictly larger than the number of neighbors of that site in $V$''.
Iterate this procedure for the new volume $V\setminus E_1$, and the new
matrix $\Delta^{V\setminus E_1}$ defined by
\begin{eqnarray*}
\Delta^{V\setminus E_1}_{x,y} &=& \Delta^V_{x,y} \ \mbox{if } x,y\in
V\setminus E_1\nonumber\\
&=& 0 \ \mbox{otherwise},
\end{eqnarray*}
and so on. If at the end some non-empty subset $V_f$ is left, $\eta$
satisfies, for all $x\in V_f$,
\[
\eta(x) \leq \sum_{y\in V_f, y\not= x} (-\Delta_{x,y}^V).
\]
The restriction $\eta|_{V_f}$ is called a {\sl forbidden
subconfiguration}. If $V_f$ is empty, the configuration is called {\sl
allowed}, and the set $\mathcal{A}_V$ of allowed configurations satisfies

\begin{proposition}
\[
\mathcal{A}_V=\re_V.
\]
\end{proposition}

The main ingredient to prove this result (see \cite{IP}, \cite{Speer}) is
the fact that toppling or adding cannot create a forbidden
subconfiguration. The set $\aaa_V$ is thus closed under the dynamics and
contains the maximal configuration $\eta_{max}$.

Remark that the burning algorithm implies that for $V'\supset V$, and
$\eta\in\Omega_V$, $\eta\not\in \re_V$, then any $\zeta \in \Omega_{V'}$
such that $\zeta|_V = \eta$ satisfies $\zeta \not\in \re_{V'}$. Indeed,
the property of having a forbidden subconfiguration in $V_f$ only depends
on the heights at sites $x\in V_f$. Therefore $\eta\in \re_{V'}$ implies
$\eta|_V \in \re_V$. This ``consistency" property will enable us to define
allowed configurations on infinite sets.

\subsection{Expected Toppling Numbers}\label{topnumber}
For $x,y\in V$ and $\eta\in \Omega_V$, let $n_V(x,y,\eta)$ denote the
{\sl number of topplings} at site $y\in V$ by adding a grain at $x\in V$,
i.e. the number of times we have to apply the operator $T_{\Delta^V}
(\cdot,y)$ to relax $\eta^x$. Define
\begin{equation}\label{top}
G_V (x,y) = \int \mu_V(d\eta )\ n_V(x,y,\eta ).
\end{equation}
Writing down balance between inflow and outflow at site $y$, one obtains
(cf. \cite{Dhar})
\[
\sum_{z\in V}\Delta_{x,z}^V G_V(z,y) = \delta_{x,y},
\]
which yields
\[
G_V(x,y) = (\Delta^V)^{-1}_{x,y}.
\]
In the limit $V\uparrow S$ (where $S$ is $\Z^d$ or the infinite tree),
$G_V$ converges  to the Green's function of the simple random walk on $S$.

\subsection{ Some specific results for the tree}\label{spetree}
When $V_n$ is a binary tree of $n$ generations,  many explicit results
have been obtained in \cite{DharMajum}. We summarize here the results we
need for the construction in infinite volume.
\begin{enumerate}
\item
When adding a grain on a particular site $0\in V_n$ of height $3$, the
set of toppled sites is the connected cluster $C_3(0,\eta)$ of sites
including $0$ having height $3$. This cluster is distributed as a {\sl
random animal} (i.e. its distribution only depends on its cardinality,
not on its form). Moreover
\begin{equation}\label{asymptotisch}
\lim_{n\uparrow\infty} \mu_{V_n} \left (|C_3 (0,\eta ) | = k \right)
\simeq Ck^{-3/2}
\end{equation}
as $k$ goes to infinity. The notation $\simeq$ means that if we multiply
the left hand side of (\ref{asymptotisch}) by $k^{3/2}$, then the limit
$k\rightarrow\infty$ is some strictly positive constant $C$.
\item
When adding a grain on site $x$, the expected number of topplings
 at site $y$ satisfies
\begin{equation}\label{limtop}
\lim_{n\uparrow\infty} \int \mu_{V_n} (d\eta ) \ n_{V_n}(x,y,\eta ) =
G(x,y),
\end{equation}
where $G(x,y)$ is the Green's function of the simple random walk on the
infinite tree, i.e.
\begin{equation}\label{green}
G(0,x) = C2^{-|x|},
\end{equation}
and $|x|$ is the ``generation number'' of $x$ in the tree.
\item
The correlations in the finite volume measures $\mu_{V_k}$ can be
estimated in terms of the eigenvalues of a product of {\sl transfer
matrices}. This formalism is explained in detail in \cite{DharMajum},
section 5: Let $f,g$ be two local functions whose dependence sets (see
below a precise definition) are separated by $n$ generations. To estimate
the truncated correlation function
\begin{equation}\label{trunc}
\mu_{V_k} \left( f;g\right) = \int fg d\mu_{V_k}  - \int f d\mu_{V_k}
\int g d\mu_{V_k},
\end{equation}
consider the product of matrices
\begin{equation}
M^k_n=\prod_{i=1}^n \left( \begin{array}{cc}
          1+\gamma^{k,n}_i & 1+\gamma^{k,n}_i\\
      1 & 2+\gamma^{k,n}_i
      \end{array} \right),
\end{equation}
where $\gamma^{k,n}_i\in [0,1]$. Let $\lambda^{n,k}_m$ (resp.
$\lambda^{n,k}_M$) denote the smallest (resp. largest) eigenvalue of
$M^k_n$. Then
\begin{equation}
\mu_{V_k} \left( f;g \right) \leq C(f,g)
\frac{\lambda^{n,k}_m}{\lambda^{n,k}_M}.
\end{equation}
If $f$ and $g$ have dependence sets ``deep within" $V_k$, then
$\gamma^{k,n}_i$ is very close to one, and the correlations are governed
by the maximal and minimal eigenvalues of $M_n= \left( \begin{array}{cc}
          2 & 2\\
      1 & 3
      \end{array} \right)^n$.
\end{enumerate}

\section{Main results}
\subsection{Notation, definitions}
From now on, $S$ denotes the infinite rootless binary tree, $V\subset S$
a finite subset of $S$; $\Omega_V$ is the set of stable configurations in
$V$, i.e. $\Omega_V = \{ \eta: V\rightarrow \{1,2,3 \} \}$, and the set
of all infinite volume stable configurations is $\Omega = \{ 1,2,3 \}^S$.
The set $\Omega$ is endowed with the product topology, making it into a
compact metric space. For $\eta\in \Omega$, $\eta|_V$ is its restriction
to $V$, and for $\eta,\zeta\in\Omega$,  $\eta_V\zeta_{V^c}$ denotes the
configuration whose restriction to $V$ (resp. $V^c$) coincides with
$\eta|_V$ (resp. $\zeta|_{V^c}$). As in the previous section,
$\re_V\subset \Omega_V$ is the set of all allowed (or recurrent)
configurations in $V$, and we define
\begin{equation}\label{allowed}
\re=\{\eta\in\Omega:\forall V\subset S \ \mbox{finite,} \ \eta|_V\in
\re_V\}.
\end{equation}
A function $f:\Omega\rightarrow \R$ is {\sl local} if there is a finite
$V\subset S$ such that $\eta|_V = \zeta|_V$ implies $f(\eta ) = f(\zeta)$.
The minimal (in the sense of set ordering) such $V$ is called {\sl
dependence set} of $f$ and is denoted by $D_f$. A local function can be
seen as a function on $\Omega_V$ for all $V\supset D_f$ and every
function on $\Omega_V$ can be seen as a local function on $\Omega$. The
set $\mathcal{L}$ of all local functions is uniformly dense in the set
$\mathcal{C} (\Omega)$ of all continuous functions on $\Omega$.

All along the paper, we use the following notion of limit by inclusion
for a function $f$ on the finite subsets of the tree with values in a
metric space $(K,d)$:

\begin{definition}
Let $\mathcal{S}=\{V\subset S,\,V\,\,\mbox{finite}\}$, and
$f:\mathcal{S}\rightarrow(K,d)$. Then
\[
\lim_{V\uparrow S}f(V)=\kappa
\]
if for all $\epsilon>0$, there exists $V_0\in S$ such that for all
$V\supset V_0$, $d(f(V),\kappa)<\epsilon$.
\end{definition}

\begin{definition}
A collection of probability measures $\nu_V$ on $\Omega_V$ is a Cauchy
net if for any local $f$ and for any $\epsilon>0$ there exists $V_0\supset
D_f$ such that for any $V,V'\supset V_0$
\[
|\int f(\eta ) \nu_V (d\eta ) - \int f(\eta ) \nu_{V'} (d\eta ) |\leq
\epsilon.
\]
\end{definition}

A Cauchy net converges to a probability measure $\nu$ in the following
sense: The mapping
\[
\Psi :\mathcal{L}\rightarrow \R;\ f\mapsto \Psi (f) = \lim_{V\uparrow S}
\int f d\nu_V
\]
defines a continuous linear functional on $\mathcal{L}$ (hence on
$\mathcal{C}(\Omega )$) which satisfies $\Psi(f)\geq 0$ for $f\geq 0$ and
$\Psi (1) = 1$. Thus by Riesz representation theorem there exists a
unique probability measure on $\Omega$ such that $\Psi (f) = \int fd\nu$.
We denote $\nu_V\rightarrow\nu$, and call this $\nu$ the {\sl infinite
volume limit} of $\nu_V$.

We will also often consider an enumeration of the tree $S$, $\{x_0, x_1,
\ldots,x_n,\ldots\}$, and put
\begin{equation}\label{subtree}
T_n=\{ x_0,\ldots , x_n\}.
\end{equation}

\subsection{ Thermodynamic limit of stationary measures}

\begin{theorem}\label{perfect}
The set $\re$ defined in (\ref{allowed}) is an uncountable perfect set,
i.e.
\begin{enumerate}
\item $\re$ is compact,
\item The interior of $\re$ is empty,
\item For all $\eta\in\re$ there exists a sequence $\eta_n\not= \eta$,
$\eta_n\in\re$, converging to $\eta$.
\end{enumerate}
\end{theorem}

For $\eta\in\Omega$, we denote by $C_3(0,\eta)$ the nearest neighbor
connected cluster of sites containing the origin and having height 3.

\begin{theorem}\label{thermo}
The finite volume stationary measures $\mu_V$ defined in
(\ref{finitevolume}) form a Cauchy net. Their infinite volume limit $\mu$
satisfies
\begin{enumerate}
\item $\mu(\re )=1$,
\item $\mu$ is translation invariant and exponentially mixing,
\item $\mu \left( \eta: |C_3(0,\eta )| <\infty  \right)=1$,
\item $\int |C_3 (0,\eta )| \mu (d\eta ) = \infty$.
\end{enumerate}
\end{theorem}

\noindent {\bf Remark:}  Point 3. above remains true for the set
$C_1(0,\eta)$, the nearest neighbor connected cluster of sites containing
the origin and having height 1, and probably also for $C_2(0,\eta)$ but
this we have not been able to prove.

\subsection{Infinite volume dynamics}
The finite volume addition operators $a_{x,V}$ (cf. (\ref{axV})) can be
extended to $\Omega$ via
\begin{equation}\label{axv}
a_{x,V}:\Omega\rightarrow\Omega;\eta\mapsto a_{x,V}\eta =
(a_{x,V}\eta|_V)_V\eta_{V^c}.
\end{equation}

\begin{proposition}\label{a_x}
\noindent
\begin{enumerate}
\item
There exists a subset $\Omega'$ of $\re$ with $\mu(\Omega')=1$ on which
the limit
\begin{equation}\label{iva}
\lim_{V\uparrow S} a_{x,V} \eta = a_x \eta
\end{equation}
exists, and $a_x\eta\in\Omega'$.
\item
The measure $\mu$ of theorem \ref{thermo} is invariant under the action
of $a_x$.
\item
For every $\eta\in\Omega'$, $a_x(a_y\eta)=a_y(a_x\eta)$, for all $x,y\in
S$.
\end{enumerate}
\end{proposition}

Part 2 implies that the infinite volume addition operators $a_x$ (cf.
(\ref{iva})) define norm 1 operators on $L^p(\mu )$, for $1\leq p\leq
\infty$ via
\[
(a_x f)(\eta ) = f(a_x\eta ).
\]

We now construct a Markov process on $\mu$-typical infinite volume
configurations which can be described intuitively as follows. Let
$\vi:S\rightarrow (0,\infty)$; this function will be the {\sl addition
rate function}. To each site $x\in S$ we associate a Poisson proces $\na$
(for different sites these Poisson processes are mutually independent)
with rate $\vi (x)$. At the event times of  $\na$ we ``add a grain" at
$x$, i.e. we apply the addition operator $a_x$ to the configuration. Then
$L_V^\vi$ introduced in (\ref{2.10}) generates a pure jump Markov process
on $\Omega$. Indeed, this operator is well-defined and bounded on any
$L^p (\mu )$ space by Proposition \ref{a_x}, which implies

\begin{proposition}\label{prop4.2}
$L_V^{\vi}$ is the $L^p(\mu)$ generator of the stationary Markov process
defined by
\[
\exp(tL_V^\vi)f=\int \left(\prod_{x\in V} a_x^{\na} f\right) d\pee,
\]
where $\pee$ denotes the joint distribution of the independent Poisson
processes $\na$, and $f\in L^p(\mu)$.
\end{proposition}

The following condition on the addition rate $\vi$ is crucial in our
construction. Remember $|x|$ is the generation number of $x$:\\

\begin{equation}\label{conC}
\mbox{\bf Summability Condition:}\;\;\;\qquad\qquad\qquad\sum_{x\in
S}\vi(x) 2^{-|x|}<\infty
\end{equation}

This condition ensures that the number of topplings at any site $x\in S$
remains finite during the addition process.

\begin{theorem}\label{process}
If $\vi $ satisfies condition (\ref{conC}), then we have
\begin{enumerate}
\item
The semigroups $S_V^{\vi} (t) = \exp (tL_V^{\vi})$ converge strongly in
$L^1 (\mu)$ to a semigroup $S_\vi(t)$.
\item
$S_\vi (t)$ is the $L^1 (\mu )$ semigroup of a stationary Markov process
$\{ \eta_t:t\geq 0 \}$ on $\Omega$.
\item
For any $f\in \mathcal{L}$,
\[
\lim_{t\downarrow 0}\frac{S_\vi(t)f -f}{t}=L^{\vi}f=\sum_{x\in
S}\vi(x)[a_x f-f],
\]
where the limit is taken in $L^1(\mu)$.
\end{enumerate}
\end{theorem}

\noindent
{\bf Remarks.}\\
1. In Proposition \ref{5.5}, we show that $S_{\vi}(t)$ is a strongly
continuous function of $\vi$.

\noindent 2. In Proposition \ref{5.6}, we show that condition (\ref{conC})
is in some sense optimal.

\begin{theorem}\label{cadlag}
The process $\{\eta_t:t\geq 0\}$ of Theorem \ref{process} admits a cadlag
version (right-continuous with left limits).
\end{theorem}

The intuitive description of the process $\{ \eta_t:t\geq 0 \}$ is
actually correct under condition ({\ref{conC}), i.e. the process has a
representation in terms of Poisson processes:

\begin{theorem}\label{poissonrep}
If $\vi$ satisfies condition (\ref{conC}), for $\mu\times\pee$ almost
every $(\eta, \omega)$ the limit
\[
\lim_{V\uparrow S} \prod_{x\in V} a_x^{\na (\omega)} \eta = \eta_t
\]
exists. The process $\{\eta_t :t\geq 0\}$ is a version of the process of
Theorem \ref{process}, i.e. its $L^1(\mu)$ semigroup coincides with
$S_\vi(t)$.
\end{theorem}

Finally, we can slightly generalize Theorem \ref{poissonrep} in order to
define the dynamics starting from a measure stochastically below $\mu$.
For $\eta,\zeta\in \Omega$, $\eta\leq\zeta$ if for all $x\in S$,
$\eta(x)\leq \zeta(x)$. A function $f:\Omega\rightarrow\R$ is {\sl
monotone} if $\eta\leq\zeta$ implies $f(\eta )\leq f(\zeta)$. Two
probability measures $\mu$ and $\nu$ satisfy $\mu\leq \nu$ if for all
monotone functions, $\int fd\mu\leq \int f d\nu$.

\begin{theorem}\label{extended}
Let $\mu'\leq \mu$. If $\vi$ satisfies condition (\ref{conC}), for
$\mu'\times\pee$ almost every $(\eta, \omega)$ the limit
\[
\lim_{V\uparrow S} \prod_{x\in V} a_x^{\na(\omega)} \eta = \eta_t
\]
exists. The process $\{\eta_t :t\geq 0\}$ is Markovian with $\eta_0$
distributed according to $\mu'$.
\end{theorem}

\noindent {\bf Remark.} The last Theorem implies  that $\eta\equiv 1$ can
be taken as initial configuration.

\section{Proofs}
This section is devoted to the proofs of the results described above.
Some of them will be put in a slightly more general framework so that they
can be applied to other cases (where $S$ is not a binary tree or where we
have other addition operators $a_x$) as soon as the existence of a
thermodynamic limit of the finite volume stationary measures is
guaranteed. The essential cause of difficulty is the non-locality of the
addition operators. The essential simplification is the abelian property
which enables us to
think of the $a_x$ as complex numbers of modulus one.\\

\subsection{ Thermodynamic limit of stationary measures}

\noindent{\bf Proof of Theorem \ref{perfect}:}\\
1,2. If $\eta\in \re$ and $\zeta\geq\eta$, then $\zeta\in\re$ (by the
burning algorithm $\zeta|_V\geq \eta|_V$ implies that $\zeta|_V\in
\re_V$). Since $\eta\equiv 2$ is in $\re$ (again by the burning
algorithm), we conclude that $\re$ is uncountable. To see that $\re$ has
empty interior, notice that if $\eta\in\re$, there does not exist $x,y\in
S$ nearest neighbors such that $\eta (x)=\eta(y)=1$ (that way,
$\eta|_{\{x,y\}}$ would be a forbidden subconfiguration). Finally $\re$
is closed as intersection of closed sets.

3. Let  $\eta_{max}$ be the maximal configuration, $\eta_{max} (x)=3$ for
all $x\in S$. If $\eta|_V\in \re_V$, then
$\eta_V(\eta_{max})_{V^c}\in\re$. Therefore any $\eta\in \re$ containing
an infinite number of sites $x$ for which $\eta(x)\not= 3$ has property 3
of Theorem \ref{perfect}. If $\eta\in\re$ contains only a finite number
of sites having height $1$ or 2, then we choose a sequence $\Sigma=\{ x_n
:n\in\N \}\subset \{x\in S:\eta(x)=3 \ \mbox{and}\ \eta(y)=3 \ \mbox{for
any neighbor of}\ x \}$ such that two elements of $\Sigma$ are never
nearest neighbors, and $|x_n|$ is strictly increasing in $n$. We then
define $\eta_n (x) = \eta (x)$ for $x\in S\setminus\ \{ x_k\in\Sigma:
0\leq k\leq n \}$ and $\eta_n (x_k )=2$ for $x_k\in\Sigma,\,k\geq n+1$.
These $\eta_n$ belong to $\re$ by the burning algorithm, and
$\eta_n\rightarrow\eta$. \QED

\begin{flushleft}
{\bf Proof of Theorem  \ref{thermo}:}
\end{flushleft}
We use $T_n$ introduced in (\ref{subtree}), but with the $x_i$ such that
$n\leq m$ implies that the generation numbers satisfy $|x_n| \leq |x_m|$.
Then we have
\begin{equation}\label{log}
|x_n|\simeq \log_2 n.
\end{equation}
To prove that the probability measures $\mu_{V}$ form a Cauchy net, it is
sufficient to show that for any local function $f:\Omega\rightarrow\R$ we
have
\begin{equation}\label{sum}
\sum_{n}|\int f d\mu_{T_n} - fd\mu_{T_{n+1}}| <\infty.
\end{equation}
We do it for $f(\eta ) = \eta (x_0 )$ (a general local function can be
treated in the same way), by giving an upper bound of the difference
$\int f d\mu_{T_n}-\int fd\mu_{T_{n+1}}$ by a truncated correlation
function (cf. (\ref{trunc})). Then we estimate the latter by the transfer
matrix method (cf. Section \ref{spetree}, part 3). We abbreviate in what
follows $\mu_n=\mu_{T_n}$, $\re_n=\re_{T_n}$.

\begin{lemma}
\[
|\mu_{n+1} [\eta(x_0)] -\mu_n [\eta(x_0)]| \leq C\mu_{n+1} [ \eta(x_0); I
(\eta(x_{n+1})=3)]
\]
\end{lemma}

\noindent{\bf Proof:} By the burning algorithm, every $\eta\in\re_n$ can
be extended to an element of $\re_{n+1}$ by putting $\eta(x_{n+1})=3$.
Moreover
\[
\{\eta|_{T_n}:\eta\in\re_{n+1},\ \eta (x_{n+1})=3 \} = \re_n,
\]
thus
\begin{equation}\label{draai}
\mu_{n+1} [ \eta (x_{n+1} )=3 ] = \frac{|\re_n |}{|\re_{n+1}|},
\end{equation}
which yields
\begin{eqnarray*}
\mu_n (\eta(x_0)) &=& \sum_{\eta \in \re_n}\frac{1}{|\re_n|}\eta(x_0)
\nonumber\\
&=& \sum_{\eta\in\re_{n+1}}\frac{1}{|\re_{n+1}|} \eta(x_0) I
(\eta(x_{n+1}=3)
\frac{|\re_{n+1}|}{|\re_n|}\nonumber\\
&=& \sum_{\eta\in\re_{n+1}} \frac{1}{|\re_{n+1}|} \eta(x_0) I(\eta
(x_{n+1})=3 ) \frac{1}{\mu_{n+1} [\eta(x_{n+1})=3]}.
\end{eqnarray*}
Therefore
\[
|\mu_{n+1}(\eta(x_0))-\mu_n(\eta(x_0))| \leq \frac{\mu_{n+1}
[\eta(x_0);I(\eta(x_{n+1})=3)]}{\mu_{n+1} [\eta(x_n+1)=3]}
\]
The Lemma follows now from (\ref{draai}), and the fact that $|\re_n|$
grows like $e^{cn}$ for some $c\geq \log 2$. \QED

Recalling Section \ref{spetree}, part 3, we have
\begin{equation}
\mu_n [\eta(x_0); I(\eta(x_n)=3)] \leq C \frac
{\lambda_m^{|x_n|,|x_n|}}{\lambda_M^{|x_n|,|x_n|}}
\end{equation}

\begin{lemma}
\[
\sum_{n=1}^{+\infty}\frac{\lambda_m^{|x_n|,|x_n|}}{\lambda_M^{|x_n|,|x_n|}}
<+\infty
\]
\end{lemma}

\noindent{\bf Proof:} We abbreviate
$\lambda_m^{(n)}=\lambda_m^{|x_n|,|x_n|},\,
\lambda_M^{(n)}=\lambda_M^{|x_n|,|x_n|},\,M(n)=M^n_{|x_n|},\,
\gamma_i=\gamma_i^{|x_n|,|x_n|}$. Remember $0\leq
\gamma_i^{|x_n|,|x_n|}\leq 1$ is close to one for $i\ll n$ and $n$ large.
In terms of the trace and the determinant of $M(n)$ we have
\begin{eqnarray*}
\lambda_M^{(n)} &=& \frac{1}{2} \left( \mbox{Tr}(M(n)) +
\sqrt{ [\mbox{Tr}(M(n))]^2-4\mbox{det}(M(n))}\right)\nonumber\\
\lambda_m^{(n)} &=& \frac{1}{2} \left( \mbox{Tr}(M(n)) - \sqrt{
[\mbox{Tr}(M(n))]^2-4\mbox{det}(M(n))}\right).
\end{eqnarray*}
Therefore,
\[
\lim_{n\uparrow\infty}
\left(\frac{\lambda_m^{(n)}}{\lambda_M^{(n)}}\right)
\left(\frac{[\mbox{Tr}(M(n))]^2}{\mbox{det} (M(n))}\right) =1.
\]
To prove the Lemma we show that (cf. (\ref{log}))
\begin{equation}
\left(\frac{\mbox{det} (M(n))}{[\mbox{Tr}(M(n))]^2}\right)\leq
(\frac{4}{9})^{|x_n|}.
\end{equation}
Use
\[
\mbox{det}(M(n)) = \prod_{i=1}^{|x_n|} (1+\gamma_i)^2,
\]
and
\begin{eqnarray*}
\mbox{Tr}(M(n)) &\geq & \mbox{Tr}\left(\prod_{i=1}^{|x_n|} \left(
\begin{array}{cc}
          1+\gamma_i & 0\\
      0 & 2+\gamma_i
      \end{array} \right)\right)\nonumber\\
&=& \prod_{i=1}^{|x_n|} (1+\gamma_i) + \prod_{i=1}^{|x_n|}
(2+\gamma_i),\nonumber
\end{eqnarray*}
to estimate (for $1\leq i\leq |x_n|,\,2(2+\gamma_i)\geq 3(1+\gamma_i)$)
\begin{eqnarray*}
\left(\frac{\mbox{det} (M(n))}{[\mbox{Tr}(M(n))]^2}\right)&\leq &
\left(1+2\prod_{i=1}^{|x_n|}\left[\frac{2+\gamma_i}{1+\gamma_i}\right] +
\prod_{i=1}^{|x_n|}
\left[\frac{2+\gamma_i}{1+\gamma_i}\right]^2\right)^{-1}\nonumber\\
&\leq & (1+2.(3/2)^{|x_n|} + (3/2)^{2|x_n|})^{-1}\nonumber\\
&\leq & (4/9)^{|x_n|}.
\end{eqnarray*}

For a general local function $f$, we have to replace $|x_n|$ by
$|x_n|-N_0$, where $N_0$ is the number of generations involved in the
dependence set of $f$. Since $f$ is local, $N_0$ is finite, hence the
convergence in (\ref{sum}) is unaffected.\QED

\subsection{Infinite volume toppling operators}

\begin{definition}\label{normaldef}
Given the finite volume addition operators $a_{x,V}$ (defined in
(\ref{axv})) acting on $\Omega$, we call a configuration $\eta\in\Omega$
{\it normal} if for every $x\in S$ there exists a minimal finite set
$V_x(\eta)\subset S$ such that for all $V'\supset V\supset V_x(\eta)$
\[
a_{x,V'}\eta = a_{x,V} \eta.
\]
\end{definition}

In other words, for a normal $\eta$, outside $V_x(\eta)$, no sites are
affected by adding a grain at $x$. In our case, when a particle is added
at some site $x\in S$, the cluster of toppled sites coincides with the
cluster $C_3(x,\eta)$ of sites having height $3$ including $x$, thus
\begin{equation}\label{Vx}
V_x(\eta)=C_3(x,\eta )\cup \partial_e C_3(x,\eta),
\end{equation}
where $\partial_e$ denotes the exterior boundary. Notice that for a
normal configuration $\eta$, by definition,
\begin{equation}
a_x(\eta)=\lim_{V\uparrow S}a_{x,V}(\eta)=a_{x,V_x(\eta)}(\eta)
\end{equation}\\

\noindent{\bf Proof of Proposition \ref{a_x}:}\\
1. We show that there is a full measure set $\Omega'$ of normal
configurations. From (\ref{asymptotisch}) and Theorem \ref{thermo},
\[
\int I(|C_3(x,\eta)|=n)d\mu \simeq Cn^{-3/2}.
\]
Therefore $\mu$ concentrates on the set $\Omega'$ of configurations for
which all the clusters $C_3 (x, \eta)$ are finite, hence for which $\eta$
is normal. Moreover this set $\Omega'$ is closed under the action of the
addition operators $a_y$, since (cf. (\ref{Vx}))
\begin{equation}\label{C3xy}
C_3 (x ,a_y\eta ) \subset V_x(\eta ) \cup V_y(\eta).
\end{equation}
\noindent 2. Choose $\epsilon >0$, pick a local function $f$, fix
$V_n\uparrow S$ and $n_0$ such that $n\geq n_0$ implies
\begin{equation}
\mu  \{ \eta\in\Omega: V_x (\eta )\not\subset V_n \} \leq
\frac{\epsilon}{4\| f \|_{\infty} +1}.
\end{equation}
This $n_0$ exists since $\mu$ concentrates on normal configurations. We
estimate
\begin{eqnarray*}
\big|\int f(a_x\eta ) d\mu - f(\eta ) d\mu \big|
&\leq & \big| \int f(a_{x,V_n}\eta ) d\mu - \int f(\eta ) d\mu \big|\nonumber\\
& +& 2 \| f \|_\infty \mu \{\eta\in\Omega:V_x(\eta ) \not\subset V_n \}
\nonumber\\
&\leq & \lim_m \big| \int f(a_{x,V_n}\eta ) d\mu_{V_m} - \int f(\eta )
d\mu_{V_m} \big|
+\frac{\epsilon}{2}\nonumber\\
&\leq & \frac{\epsilon}{2} + 2 \| f \|_\infty \lim_{m} \mu_{V_m}
\left(a_{x,V_n} (\eta ) \not= a_{x,V_m} (\eta )\right)\nonumber\\
&=& \frac{\epsilon}{2} + 2 \|f \|_\infty
\left( 1- \lim_m \mu_{V_m} (V_x(\eta )\subset V_n) \right)\nonumber\\
&=& \frac{\epsilon}{2} + 2 \|f \|_\infty \left( 1-\mu ( V_x (\eta )
\subset V_n ) \right)\leq \epsilon.
\end{eqnarray*}
In the last step we used that the indicator $I(V_x (\eta )\subset V_n)$
is a local function.

3. Let $\eta\in\Omega'$, $x,y\in S$ be two different sites and $V\supset
V_x(\eta ) \cup V_x(a_x\eta )\cup V_x (a_y \eta )$. Since $a_{x,V}$ and
$a_{y,V}$ commute, we have
\begin{eqnarray*}
& &a_x (a_y \eta) = a_x (a_{y,V} \eta)=a_{x,V} (a_{y,V} \eta )\nonumber\\
&=& a_{y,V} (a_{x,V} \eta ) =a_{y,V} (a_x\eta ) =a_y (a_x\eta ).
\end{eqnarray*}
\QED

\subsection{Infinite volume semigroup}
We now turn to the proofs of Theorems \ref{process} and \ref{cadlag}.

\begin{definition} We define
the cluster of $\eta\in\Omega$ at $x\in S$ as
\begin{equation}\label{cluster}
 \ce (x,\eta ) = \{ y\in S: a_y \eta (x) \not= \eta (x) \},
\end{equation}
and put
\begin{equation}\label{G}
G_\mu (x,y) = \int I(y\in \ce(x,\eta )) d\mu (\eta ).
\end{equation}
Finally for $\vi :S\rightarrow [0,\infty)$, write
\[
\|f\|_\vi=\sum_{x\in S}\vi(x)\int\mu(d\eta)|f(a_x\eta)-f(\eta)|,
\]
\[
\bee_\vi=\{f:\Omega\rightarrow \R: f \ \mbox{bounded},\|f\|_\vi<\infty \}.
\]
\end{definition}

\begin{lemma}\label{prop3}
If
\begin{equation}\label{geemu}
\sum_{x\in S} \vi (x) G_\mu (y,x) <\infty\ \mbox{for all}\ y\in S,
\end{equation}
then all local functions are in $\bee_\vi$.
\end{lemma}

\noindent{\bf Proof:} Let $f$ be a local function with dependence set
$D_f$. Then $f(a_x\eta)\not= f(\eta)$ if for $y\in D_f$, $a_x\eta(y)\not=
\eta(y)$, i.e. $x\in\ce(y,\eta)$:
\begin{eqnarray*}
\| f\|_\vi &=& \sum_{x\in S} \vi (x)\int |a_x f -f | d\mu\nonumber\\
&=& \int\sum_{x\in \cup_{y\in D_f} \ce(y,\eta )}\vi (x) |a_x f- f| d\mu
\nonumber\\
&\leq & 2 \|f \|_\infty\sum_{x\in S} \vi (x)\int I(x\in \cup_{y\in D_f}
\ce(y,\eta ))d\mu
\nonumber\\
&\leq & 2\|f\|_\infty\sum_{y\in D_f}\sum_{x\in S} G_\mu (y,x) \vi (x)
<\infty.
\end{eqnarray*}
\QED

The next lemma provides a link between $G_\mu$ and the Green's function
for simple random walk on $S$, i.e., between conditions (\ref{geemu}) and
(\ref{conC}).

\begin{lemma}\label{lemma5.4}
\[
G_\mu (x,y) \leq \sum_{z\sim x} G(y,z)=\delta_{x,y} + 3G(x,y),
\]
where $z\sim x$ means that $z$ and $x$ are neighbors.
\end{lemma}

\noindent{\bf Proof:} We have to estimate the probability that $a_x \eta
(y) \not= \eta (y)$. If by adding a grain at $x$ we influence $y$, this
can only be achieved by the toppling of one of the nearest neighbor sites
of $y$. Since $\mu$ concentrates on normal configurations,
\begin{eqnarray}\label{nlim}
\mu \left(a_x\eta (y) \not= \eta (y)\right)&=& \lim_{V\uparrow S} \mu
\left(a_x \eta (y) \not=\eta (y), V_x (\eta )\cup V_y (\eta )\subset V
\right)\nonumber\\
&=&\lim_{V\uparrow S}\lim_{W\uparrow S}\mu_W \left(a_{x,V} \eta (y)
\not=\eta (y), V_x (\eta )\cup V_y (\eta )\subset V
\right)\nonumber\\
&=& \lim_{V\uparrow S}\lim_{W\uparrow S}\mu_W \left(a_{x,W} \eta (y)
\not=\eta (y), V_x (\eta )\cup V_y (\eta )\subset V
\right)\nonumber\\
&\leq & \lim_{W\uparrow S}\mu_W \left(a_{x,W} \eta (y) \not=\eta (y)
\right)\nonumber\\
&\leq & \lim_{W\uparrow S} \mu_W \left(\exists \ z\in W, z\sim y, n_W
(z,y,\eta)\geq 1\right)
\nonumber\\
&\leq & \lim_{W\uparrow S} \sum_{z\sim y} \int d\mu_W (\eta ) n_W
(z,y,\eta )\nonumber\\
&=&\sum_{z\sim y} G (z,y),
\end{eqnarray}
where we used (\ref{top}),(\ref{limtop}), (\ref{Vx}) and (\ref{C3xy}).
\QED

The following Lemma finishes the proof of Theorem \ref{process} and shows
that $\bee_\vi$ is a natural core for the domain of the generator of the
infinite volume semigroup.

\begin{lemma}\label{lemma2}
\noindent
\begin{enumerate}
\item
For $f\in\bee_\vi$ the net
\begin{equation}
S^\vi_V (t) f =\exp (t L^\vi_V) f= \exp \left(t\sum_{x\in
V}\vi(x)(a_x-I)\right)f
\end{equation}
converges in $L^1(\mu )$ (as $V\uparrow S$) to a function $S_\vi(t) f\in
L^1 (\mu )$. $f\mapsto S_\vi(t) f$ defines a semigroup on $\bee_\vi$
which is a contraction in both $L^1(\mu )$ and $\bee_\vi$ norms.
\item
Under condition (\ref{geemu}), the semigroup $S_\vi(t)$ corresponds to a
unique Markov process on $\Omega$.
\end{enumerate}
\end{lemma}

\noindent{\bf Proof:} We denote by $\|f\|$ the $L^1(\mu)$-norm of $f$,
and we abbreviate $S_V(t)=S_V^\vi(t),\,S(t)=S^\vi(t),\,L_V=L_V^\vi$.

\noindent 1. First note that $S_V (t)$ is well-defined on $L^1(\mu )$ by
Proposition \ref{prop4.2}. By the abelian property (Proposition
\ref{a_x}, part 3) we can write for $V\subset V' \subset S$:
\[
\| S_V(t)f - S_{V'}(t)f \| = \| (S_{V'\setminus V}(t)-I) S_V(t) f \|
\]
By Proposition \ref{prop4.2}, $S_V (t)$ is the semigroup of a stationary
Markov process and hence a contraction on $L^1(\mu)$. Therefore
\begin{eqnarray}\label{therefore}
&&\| S_V (t) (S_{V'\setminus V} (t) - I ) f \| \leq
\| (S_{V'\setminus V} (t) - I) f \|\nonumber\\
&=& \| \int_0^t  L_{V'\setminus V} S_{V'\setminus V} (s) f ds\|
\nonumber\\
&\leq &
\int_0^t \| L_{V'\setminus V} f \|ds\nonumber\\
&\leq & t \sum_{x\in V'\setminus V} \vi (x) \int | (a_x - I ) f | d\mu
\rightarrow 0 \ \mbox{as} \ V,V'\uparrow S,
\end{eqnarray}
where the last step follows from $f\in \bee_\vi$. Hence $S_V (t) f
\rightarrow S(t) f$ in $L^1 (\mu )$. We show that $S(t) f\in\bee_\vi$:
\begin{eqnarray*}
&&\sum_{x\in S} \vi (x)\int | S(t) f(a_x\eta ) - S(t) f(\eta ) | \mu
(d\eta )
\nonumber\\
&\leq & \sum_{x\in S} \vi (x)\int S(t)|a_x f-f| d\mu
\nonumber\\
&=& \int \sum_{x\in S} \vi (x) |a_x f -f| d\mu = \| f \|_\vi.
\end{eqnarray*}
Thus $S(t)$ is also a contraction for the $\| \cdot\|_\vi$-norm. We
finish with the semigroup property:
\begin{eqnarray*}
S(t) S(s) f &=& \lim_{V\uparrow S} S_V (t) [ S(s) f]
\nonumber\\
&=& \lim_{V\uparrow S}\lim_{W\uparrow S} S_V (t) S_W (t) f,
\end{eqnarray*}
and
\[
S(t+s)f = \lim_{V\uparrow S} S_V (t) S_V (s) f.
\]
Then, since $S_V(t)$ is a contraction in $L^1(\mu)$,
\begin{equation}\label{thencan}
\|S_V (t) S_W (t) f- S_V (t) S_V (s) f\| \leq \|S_W(s)f - S_V(s) f\|,
\end{equation}
By (\ref{therefore}), the right hand side of (\ref{thencan}) goes to zero
as $V,W\uparrow S$.

\noindent 2. If condition (\ref{geemu}) is met, then $\bee_\vi$ contains
all local functions by Lemma \ref{prop3}. Therefore, by contractivity the
semigroup $S(t)$ on $\bee_\vi$ uniquely extends to a semigroup of
contractions on $L^1 (\mu )$. Since by Proposition \ref{prop4.2}, $S_V
(t)$ is a Markov semigroup, so is $S(t)$, i.e. $S(t)1=1$,  $S(t)f\geq 0$
if $f\geq 0 $.  Hence, by Kolmogorov's theorem there is a unique Markov
process with semigroup $S(t)$. \QED

\noindent {\bf Remark.} When $\vi \equiv 1$, condition (\ref{geemu}) is
equivalent to
\[
\sum_{x\in S}\int\mu(d\eta) I(x\in\ce(y,\eta))=\int|\ce(y,\eta)|\mu(d\eta)
<+\infty,
\]
i.e., the clusters must be integrable under $\mu$. For models which
exhibit ``self-organized criticality'', $\ce(y,\eta)$ is usually a
``finite but critical percolation cluster'', implying that
$\int |\ce(y,\eta)| d\mu =\infty$ (cf. Theorem \ref{thermo} part 4,
because $\ce(y,\eta)\supset\partial_e C_3(y,\eta)$). Therefore this
formalism breaks down for addition rate $\vi\equiv 1$.

The following Lemma proves Theorem \ref{cadlag}.

\begin{lemma}\label{cadlagprop}
Under condition (\ref{geemu}), the process $\{ \eta_t:t\geq 0 \}$ of
Theorem \ref{process} is almost surely right-continuous, i.e.
\begin{equation}\label{cad}
\pee_\mu\left[ \lim_{t\downarrow 0} d(\eta_t,\eta_0) \geq \epsilon\right]
=0,
\end{equation}
where $\pee_\mu$ is its path-space measure, and the distance $d$ is
defined below (in (\ref{mey})).
\end{lemma}

\noindent{\bf Proof:} Pick a function $\Psi:S\rightarrow (0,1)$ such that
\begin{equation}\label{drac}
\sum_{x\in S} \Psi (x) =1,
\end{equation}
and
\begin{equation}\label{franc}
\sum_{x,y\in S}\vi (x) G_\mu (x,y) \Psi (y) <\infty
\end{equation}
The distance
\begin{equation}\label{mey}
d(\eta,\zeta )= \sum_{x\in S} |\eta (x) -\zeta (x) | \Psi (x)
\end{equation}
generates the product topology. Denote by $\E_\mu$ the expectation w.r.t.
$\pee_\mu$. For $f_y (\eta ) = \eta (y)$,
\[
f_y(\eta_t) - f_y(\eta _0) = \int_0^t L^\vi f_y(\eta_s) ds + M^y_t,
\]
where $M^y_t$ is a centered martingale with quadratic variation
\begin{equation}\label{sh}
\E_\mu \left[(M^y_t)^2\right] = \E_\mu \left[ \int_0^t (L^\vi f_y^2
(\eta_s )-2 f_y (\eta_s )L^\vi f_y (\eta_s )) ds \right].
\end{equation}
Using stationarity of $\eta_s$ and
\[
\int d\mu |L^\vi g| \leq 2\|g\|_\infty\sum_{x\in S} \sum_{y\in D_g}\vi (x)
G_\mu (y,x ),
\]
for a local bounded function on $\Omega$, we obtain from (\ref{sh})
\[
\E_\mu \left[(M^y_t)^2\right] \leq Ct \sum_{x\in S} \vi (x) G_\mu (y,x).
\]
Now we can estimate
\begin{eqnarray*}
& &\pee_\mu \left[\exists s\leq t:\sum_{y\in S}|\eta_s (y) - \eta_0(y)|
\Psi (y) \geq \epsilon \right]
\nonumber\\
&\leq & \pee_{\mu} \left[ \int_0^t ds\sum_{y\in S} |L^\vi f_y (\eta_s)|
\Psi (y)\geq \epsilon/2 \right] + \pee_\mu \left[ \sup_{0\leq s\leq t}
\left| \sum_{y\in S} M^y_s \Psi (y) \right| \geq \epsilon/2 \right]
\nonumber\\
&\leq & (12t/\epsilon) \sum_{x,y\in S} \vi (x) G_\mu (y,x) \Psi (y) +
(2/\epsilon)^2 \E_\mu \left| \sum_{y\in S} M^y_t \Psi (y) \right|^2
\nonumber\\
&\leq & (12t/\epsilon) \sum_{x,y\in S} \vi (x) G_\mu (y,x) \Psi (y) +
(2/\epsilon)^2\E_\mu \left[\sum_{y\in S} (M^y_t)^2\Psi (y)\right]
\nonumber\\
&\leq & tC_\epsilon\sum_{x,y\in S}\vi (x) G_\mu (y,x )\Psi (y).
\end{eqnarray*}
Here we used Markov's and Doob's inequalities in the second step and the
Cauchy-Schwarz inequality combined with (\ref{drac}) in the third step.
The result (\ref{cad}) follows.\QED

\subsection{Poisson representation}
In this section we prove Theorems \ref{poissonrep} and \ref{extended}.
Intuitively it is clear from the abelian property that the process of
which we showed existence in the previous subsection can be represented
as $\prod_{x\in S}a_x^{N^{t,x}_\vi}\eta$, where $\na$ are independent
Poisson processes of intensity $\vi(x)$.

We take $T_n$ as in (\ref{subtree}). We say that the product $\prod_{x\in
S} a_x^{n_x} \eta$ exists if for every $y\in S$ there exists $N_y$ such
that for all $m,n\geq N_y$
\[
\left|\left[\prod_{x\in T_{n}} a_x^{n_x}\eta\right] (y)
-\left[\prod_{x\in T_m} a_x^{n_x}\eta\right] (y)\right|=0.
\]
This is equivalent to the convergence of the sequence $\prod_{x\in T_n}
a_x^{n_x}\eta$ in the product topology.

\begin{lemma}\label{poissonprop}
Under condition (\ref{geemu}), the product
\[
\prod_{x\in S} a_x^{\na}\eta=\tilde\eta_t
\]
exists for $\mu$-almost every realization of $\na$ and almost every
$\eta$. The process $\{\tilde\eta_t:t\geq 0 \}$ is a version of the Markov
process of Lemma \ref{lemma2}.
\end{lemma}

\noindent{\bf Proof:} Choose a realization of $\na$ such that
\begin{equation}\label{probone}
\sum_{x\in S} \na G_\mu(x,y) < \infty
\end{equation}
for every $y$. This happens with probability one by condition
(\ref{geemu}). Define for $\eta\in\Omega'$
\begin{equation}\label{etaVn}
\eta_{T_n} (t)= \prod_{x\in T_n} a_x^{\na} \eta.
\end{equation}
Under $\mu$, $\eta_{T_n}(t)$ is stationary in $n$ and $t$. We have
\begin{eqnarray*}
&&\mu\left[ \left| (\eta_{T_n}(t))(y) -(\eta_{T_{n+1}}(t))(y) \right|
\geq 1\right]
\nonumber\\
&\leq & \int \left|\left[
a^{N^{t,x_{n+1}}_\varphi}_{x_{n+1}}\eta_{T_n}(t)\right](y) -
(\eta_{T_n}(t))(y)
\right|\mu(d\eta)\nonumber\\
&=& \int\left|\left[ a^{N^{t,x_{n+1}}_\varphi}_{x_{n+1}}\eta\right](y)
- \eta(y )\right|\mu(d\eta)\nonumber\\
&\leq & \int \sum_{j=1}^{N^{t,x_{n+1}}_\varphi}
\left|\left[a^j_{x_{n+1}}\eta\right](y)
- \left[a^{j-1}_{x_{n+1}} \eta \right](y)\right|\mu (d\eta)\nonumber\\
&\leq & 6 N^{t,x_{n+1}}_\varphi G_\mu(x_{n+1},y).
\end{eqnarray*}

In the second and last steps we used the invariance of $\mu$ under $a_x$.
By the Borel Cantelli Lemma, (\ref{probone}) implies that for almost
every realization of $\na$
\[
\mu \left[ \exists n_0:\forall n\geq n_0 \ (\eta_{T_n} (t))(y) =
(\eta_{T_{n_0}} (t))(y)\right] =1.
\]
This proves $\mu$-a.s. convergence of the product. To see that
$\tilde\eta_t$ is a version of the Markov process with semigroup
$S_\vi(t)$, combine Proposition \ref{prop4.2} with Theorem \ref{process},
part 1 to get, for any local function $f$,
\[
\int d\mu \left|\int d\pee f(\tilde\eta_t)-S_\vi(t) f\right|=0.
\]
In the preceding argument we used a particular enumeration of the
countable set $S$. But changing it gives again a process with semigroup
$S_\vi(t)$. Therefore the limiting process will not depend (up to sets of
measure zero) on the chosen enumeration of $S$. \QED

\noindent{\bf Proof of Theorem \ref{extended}:}\\
For $\eta\in \Omega'$ and $y\in S$ we have the relation (remember
(\ref{etaVn}))
\begin{equation}\label{explicit}
\eta_V(t)(y) = \eta(y) +I(y\in V)\sum_{x\in V} \na -\Delta n_V^t(y),
\end{equation}
where $n_V^t (x)$, an integer valued random variable, is the number of
topplings at site $x$ in the time interval $[0,t]$, when sand is added in
$V$. For $T_n$ defined in (\ref{subtree}) we will first prove that
$n_{T_n}^t$ increases $\mu\times\pee$ almost surely to an
$L^1(\mu\times\pee)$ random variable $n^t$, interpreted as the number of
topplings in $[0,t]$ when we add grains according to $\na$. By the
abelian property the sequence $n^t_{T_k}(0)$ is increasing in $k$. The
following estimate is similar to (\ref{nlim})
\begin{eqnarray}\label{109}
(\mu\times\pee)&\left(\right.& \left.|n^t_{T_k} (0)-
n^t_{T_{k+1}}(0)|\geq \epsilon \right) =
(\mu\times\pee)\left(n^t_{x_{k+1}}(0)\geq \epsilon \right)
\nonumber\\&\leq & \frac{1}{\epsilon}\int n^t_{x_{k+1}}(0) \mu
(d\eta)\times\pee (d\omega)
\nonumber\\
&\leq & \frac{1}{\epsilon}\lim_{V\uparrow S}\int n^t_{x_{k+1}}(0)
I(V_{x_{k+1}}(\eta)\cup V_0(\eta)\subset V) \mu(d\eta)\times\pee(d\omega)
\nonumber\\
&\leq & \frac{1}{\epsilon}\lim_{W\uparrow S}\int n^t_W(x_{k+1},0,\eta)
\mu_W (d\eta)\times\pee(d\omega)
\nonumber\\
&\leq & \frac{1}{\epsilon}t\vi (x_{k+1}) G(0,x_{k+1}).
\end{eqnarray}
In the fourth line, $n^t_W(x_{k+1},0,\eta)$ denotes the number of
topplings up to time $t$ at site $0\in W$ by adding grains at site
$x_{k+1}\in W$. By the Borel Cantelli Lemma, condition (\ref{conC})
implies the a.s. convergence of $n^t_{T_k}(0)$, and analogously of every
$n^t_{T_k}(x)$. Pick $(\eta,\omega)$ such that $n^t_{T_k}(\eta,\omega)$
converges, i.e. such that $\sup_k
n^t_{T_k}(\eta,\omega)=n^t(\eta,\omega)$ is finite (indeed,
$n^t_{T_k}(\eta,\omega)$ is an integer). If $\eta'\leq\eta$, then
$n^t_{T_k}(\eta',\omega )\leq n^t_{T_k}(\eta,\omega)$ because we can
obtain $\eta$ from $\eta'$ by adding sand at sites $x\in S$ for which
$\eta'(x)<\eta(x)$ thereby increasing the number of topplings. We thus
conclude that the convergence of $n^t_{T_k}(\eta,\omega)$ implies the
convergence of $n^t_{T_k}(\eta',\omega )$ for all $\eta'\leq\eta$.

Now let $\mu'\leq \mu$ in the FKG sense. There is a coupling $\pee_{12}$
of $\mu'\times \pee$ and $\mu\times\pee$ such that
\[
\pee_{12} \left( ((\eta_1,\omega_1),(\eta_2,\omega_2)):
\omega_1=\omega_2, \eta_1\leq\eta_2\right)=1,
\]
i.e. we use the same Poisson events and couple $\mu'$ and $\mu$ according
to the optimal coupling (see \cite{Str}). Then
\begin{eqnarray*}
&&(\mu'\times\pee) \left( n_{T_k}^t(\eta,\omega) \rightarrow n^t
(\eta,\omega) \right) = \pee_{12} \left( n^t_{T_k}(\eta_1,\omega_1)
\rightarrow n^t (\eta_1, \omega_1)\right)
\nonumber\\
&=& \pee_{12} \left( n_{T_k}^t(\eta_1,\omega_1)\rightarrow n^t
(\eta_1,\omega_1), n^t_{T_k} (\eta_2,\omega_2 ) \rightarrow n^t
(\eta_2,\omega_2 ), \omega_1 = \omega_2, \eta_1\leq\eta_2 \right)
\nonumber\\
&\geq & \pee_{12} \left( n^t_{T_k}(\eta_2,\omega_2) \rightarrow n^t
(\eta_2,\omega_2 ) \right)
\nonumber\\
&=& (\mu\times\pee) \left( n^t_{T_k}(\eta,\omega )\rightarrow n^t
(\eta,\omega ) \right) =1.
\end{eqnarray*}
This shows the $\mu'\times\pee$-almost sure convergence of $n_{T_k}^t$,
hence by (\ref{explicit}) the product $\prod_{x\in
S}a_x^{\na(\omega)}\eta'$ converges $\mu'\times\pee$ almost surely. \QED

As a further result we show that the semigroup $S_\vi(t)$ is continuous
as a function of the addition rate $\vi$. We define
\[
\ell_1  = \{ \vi: S\rightarrow [0,\infty ):\|\vi\|= \sum_{x\in S} \vi (x)
G(0,x) <\infty \}.
\]
It is a complete metric space (as a closed subset of a Banach space) with
the property: If $\vi_n\in \ell_1$, $\vi_n\uparrow\vi$ (pointwise), and
$\vi\in\ell_1$ then $\vi_n\rightarrow\vi$ in $\ell_1$.

\begin{proposition}\label{5.5}
The semigroup $S_\vi (t)$ of Theorem \ref{process} is a strongly
continuous function of $\vi$, i.e. if $\vi_n\rightarrow\vi$ in $\ell_1$,
then for any local function $f$, $S_{\vi_n}(t) f\rightarrow S_\vi (t)f$.
\end{proposition}

\noindent{\bf Proof:} Let $n_\vi^t = \lim _{k\to\infty}n_{T_k}^t $ be the
number of topplings in $[0,t]$ from sand addition at rate $\vi$.  In the
proof of Theorem \ref{poissonrep} we have shown that this random variable
is $\mu\times\pee$ almost surely well defined and, after taking limits in
(\ref{explicit}), satisfies
\begin{equation}\label{limexpli}
\eta_t = \eta_0 + N_\vi^t - \Delta n_\vi^t,
\end{equation}
where $N_\vi^t=\lim_{V\uparrow S}\sum_{x\in V}\na$. Note that if
$\psi_1\leq\psi_2$, the Poisson processes $N_{\psi_1}^t$ and
$N_{\psi_2}^t$ can be coupled in such a way that for all $x\in S$,
$N_{\psi_1}^{t,x} \leq N_{\psi_2}^{t,x}$, and hence, by the abelian
property, $n_{\psi_1}^t (x) \leq n_{\psi_2}^t (x)$. Consider a coupling
of the four Poisson processes $N_{\vi}^t$, $N_{\vi\wedge\vi_n}^t$,
$N_{\vi\vee\vi_n}^t$ and $N_{\vi_n}^t$ under which the inequalities
$X_1(t)\geq X_2 (t)$, $X_2 (t) \leq X_3 (t)$, $X_3 (t)\geq X_4 (t)$, are
satisfied with probability one. Let $\tilde\pee$ denote the law of the
marginal $(X_1,X_4)$. We have, by a reasoning similar to (\ref{109}),
\begin{eqnarray}\label{hij}
\int d\mu\left(\tilde\E |n^t_{\vi_n}(0) - n^t_{\vi}(0)|\right) &\leq&
\int d\mu\left(\tilde\E\left(n^t_{\vi_n}(0)-n^t_{\vi_n\wedge\vi}(0)\right)
\right.\nonumber\\
&+& \tilde\E\left(n_{\vi\vee\vi_n}^t(0)-n_{\vi\wedge\vi_n}^t(0)\right)
\nonumber\\
&+& \left.\tilde\E\left(n_{\vi\vee\vi_n}^t(0) - n_{\vi}^t(0)\right)\right)
\nonumber\\
&\leq& t\sum_{x\in S}|\vi_n(x)-\vi(x)|G(0,x),
\end{eqnarray}
which tends to zero for $\vi_n\rightarrow\vi$ in $\ell_1$. Take now a
local function $f$, and denote $\tilde{ D_f}=D_f\cup\partial_e D_f$,
\begin{eqnarray*}
|S_{\vi_n}(t)(f)-S_{\vi}(t)(f)| &\leq & \tilde\pee\left( n_{\vi_n}^t(x)
\not= n_{\vi}^t(x)\ \mbox{for some}\
x\in \tilde{D_f} \right)\nonumber\\
&\leq & \sum_{x\in\tilde{D_f}} \tilde\E |n_{\vi}^t (x)-n_{\vi_n}^t(x)|.
\end{eqnarray*}
Combining this with (\ref{hij}) concludes the proof. \QED

One might ask whether we can go beyond condition (\ref{conC}), which
essentially guarantees that the expected number of topplings stays finite
in the addition process. In the following proposition we show that it is
impossible to keep integrable toppling numbers and ``rate 1" addition.
The relation (\ref{tee}) should be regarded as the infinitesimal version
of (\ref{limexpli}), where $\alpha(x)$ replaces the rate $\vi(x)$. We
then show that $\vi$ has to depend on $x$.

\begin{proposition}\label{5.6}
Let $\alpha: S\rightarrow \{ 0,1 \}$ be a stationary and ergodic random
field distributed according to $\nu $. Denote by $\int \alpha (0)\nu
(d\eta) = \rho $ its density. Suppose there exists a measurable
transformation $T: \{ 0, 1 \}^S\times\Omega\rightarrow\Omega$ which
satisfies the conditions
\begin{enumerate}
\item
The measure $\mu$ of Theorem \ref{thermo} is invariant under
$T(\alpha,\cdot)$ for any $\alpha$.
\item
\begin{equation}\label{tee}
T(\alpha,\eta)(x) =\eta (x) +\alpha (x) -\Delta n(\alpha,\eta,x),
\end{equation}
with $n(\alpha,\eta,.)\in L^1(\mu)$ for $\nu$ almost every $\alpha$.
\end{enumerate}
Then, $\rho =0$.
\end{proposition}

\noindent{\bf Proof:} Taking expectation over $\mu$ in (\ref{tee}) gives
\begin{equation}\label{lapla}
\Delta \Psi (\alpha,x) =\alpha (x),
\end{equation}
where $\Psi (\alpha,x ) = \int n(\alpha,\eta,x) \mu(d\eta)$. By
stationarity of $\mu$ and $\nu$, $\Psi(\alpha,x)$ is a stationary random
field. Let $(x_t:t\geq 0)$ denote continuous time simple random walk on
$S$, starting at $0$. From (\ref{lapla}),
\[
\E\Psi(\alpha,x_t ) =\Psi (\alpha,0 ) + \E\int_0^t \alpha (x_s)ds.
\]
Divide this last line by $t$ and let $t\uparrow +\infty$. As $\nu$ is
ergodic (making the last term equal to $\rho$) and as the process
$\Psi(\alpha,x_t)$ is stationary, we conclude that $\rho=0$. \QED

\noindent {\tt Adresses:}\\ C.M.: Instituut voor
    Theoretische Fysica, K.U.Leuven, Celestijnenlaan 200D, B-3001
    Leuven, Belgium - email:
    {\tt christian.maes@fys.kuleuven.ac.be}\\
F.R.: On leave from Instituut
    voor Theoretische Fysica, K.U. Leuven, Celestijnenlaan 200D, B-3001
    Leuven, Belgium - email: {\tt f.h.j.redig@tue.nl } \\
E.S.: CNRS, UMR 6085, Universit\'e de Rouen, 76821 Mont-Saint-Aignan
cedex, France.
   - email: {\tt Ellen.Saada@univ-rouen.fr}

\end{document}